\documentclass[a4paper,11pt]{article}
\usepackage{pos}

\title{The Baikal-GVD neutrino telescope as an instrument for studying Baikal water luminescence}
\ShortTitle{Baikal water luminescence}

\author[a]{V.A.~Allakhverdyan}
\author[b]{A.D.~Avrorin}
\author[b]{A.V.~Avrorin}
\author[b]{V.M.~Aynutdinov}
\author[c]{R.~Bannasch}
\author[d]{Z.~Barda\v{c}ov\'{a}}
\author[a]{I.A.~Belolaptikov}
\author[a]{I.V.~Borina}
\author[a,1]{V.B.~Brudanin}
\author[e]{N.M.~Budnev}
\author[a]{V.Y.~Dik}
\author[b]{G.V.~Domogatsky}
\author[b]{A.A.~Doroshenko}
\author*[a,d]{R.~Dvornick\'{y}}
\author[e]{A.N.~Dyachok}
\author[b]{Zh.-A.M.~Dzhilkibaev}
\author[d]{E.~Eckerov\'{a}}
\author[a]{T.V.~Elzhov}
\author[f]{L.~Fajt}
\author[g,1]{S.V.~Fialkovski}
\author[e]{A.R.~Gafarov}
\author[b]{K.V.~Golubkov}
\author[a]{N.S.~Gorshkov}
\author[e]{T.I.~Gress}
\author[a]{M.S.~Katulin}
\author[c]{K.G.~Kebkal}
\author[c]{O.G.~Kebkal}
\author[a]{E.V.~Khramov}
\author[a]{M.M.~Kolbin}
\author[a]{K.V.~Konischev}
\author[h]{K.A.~Kopa\'{n}ski}
\author[a]{A.V.~Korobchenko}
\author[b]{A.P.~Koshechkin}
\author[i]{V.A.~Kozhin}
\author[a]{M.V.~Kruglov}
\author[b]{M.K.~Kryukov}
\author[g]{V.F.~Kulepov}
\author[h]{Pa.~Malecki}
\author[a]{Y.M.~Malyshkin}
\author[b]{M.B.~Milenin}
\author[e]{R.R.~Mirgazov}
\author[a]{D.V.~Naumov}
\author[a]{V.~Nazari}
\author[h]{W.~Noga}
\author[b]{D.P.~Petukhov}
\author[a]{E.N.~Pliskovsky}
\author[j]{M.I.~Rozanov}
\author[a]{V.D.~Rushay}
\author[e]{E.V.~Ryabov}
\author[b]{G.B.~Safronov}
\author[a]{B.A.~Shaybonov}
\author[b]{M.D.~Shelepov}
\author[a,d,f]{F.~\v{S}imkovic}
\author[a]{A.E. Sirenko}
\author[i]{A.V.~Skurikhin}
\author[a]{A.G.~Solovjev}
\author[a]{M.N.~Sorokovikov}
\author[f]{I.~\v{S}tekl}
\author[b]{A.P.~Stromakov}
\author[a]{E.O.~Sushenok}
\author[b]{O.V.~Suvorova}
\author[e]{V.A.~Tabolenko}
\author[e]{B.A.~Tarashansky}
\author[a]{Y.V.~Yablokova}
\author[c]{S.A.~Yakovlev}
\author[b]{D.N.~Zaborov}

\affiliation[a]{Joint Institute for Nuclear Research, Dubna, Russia}
\affiliation[b]{Institute for Nuclear Research, Russian Academy of Sciences, Moscow, Russia}
\affiliation[c]{EvoLogics GmbH, Berlin, Germany}
\affiliation[d]{Comenius University, Bratislava, Slovakia}
\affiliation[e]{Irkutsk State University, Irkutsk, Russia}
\affiliation[f]{Czech Technical University in Prague, Prague, Czech Republic}
\affiliation[g]{Nizhny Novgorod State Technical University, Nizhny Novgorod, Russia}
\affiliation[h]{Institute of Nuclear Physics of Polish Academy of Sciences (IFJ~PAN), Krak\'{o}w, Poland}
\affiliation[i]{Skobeltsyn Institute of Nuclear Physics, Moscow State University, Moscow, Russia}
\affiliation[j]{St.~Petersburg State Marine Technical University, St.Petersburg, Russia}

\note{Deceased.}

\emailAdd{dvornicky@fmph.uniba.sk}

\abstract{
We present data on the Baikal water luminescence collected with the Baikal-GVD neutrino telescope. This three-dimensional array of photo-sensors allows the observation of time and spatial variations of the ambient light field. We report on annual increase of luminescence activity in years 2018-2020. We observed a unique event of a highly luminescent layer propagating upwards with a maximum speed of 28\,m/day for the first time. 
}

\FullConference{37$^{\rm{th}}$ International Cosmic Ray Conference (ICRC 2021)\\
    July 12th -- 23rd, 2021\\
    Online -- Berlin, Germany}

\begin{document}
\maketitle

\section{Introduction}

Lake Baikal remains a unique place for various species of animals and plants, many of which are endemic. The key role in this environment is played by hydrophysical processes of horizontal and vertical water exchange, which supply and distribute the oxygen and organic substances at different depths of the lake. The study of the hydrodynamic processes in Lake Baikal are of particular interest for earth and life sciences. These processes are on a spatial scale from a few milimeters to tens of kilometers and advance in time periods from a few seconds to a year. Beyond limnology, better understanding of the lake environment is of particular interest for the construction and maintenance of the neutrino telescope Baikal-GVD. 

The next generation neutrino telescope Baikal-GVD is placed in the southern basin of Lake Baikal about 3.6\,km from shore at a depth of 1\,366\,m.  The main goal of the experiment is the detection of high energy astrophysical neutrinos, whose sources remain still unknown. In particular, the aim is the registration of the Cherenkov radiation emitted when secondary charged particles, created in the reactions of neutrinos with surrounding medium, are passing through the deep water in Lake Baikal. The detector itself is a three-dimensional array of photo-sensitive components called optical modules (OMs). A fully independent unit called cluster consists of 288 OMs attached on 8 strings, 7 peripheral strings surrounding the central one with a radius of 60\,m. Each string carries 36 OMs with 15\,m vertical spacing. The top and the bottom OMs are located at depths of 750\,m and 1\,275\,m, respectively. In 2016, the first cluster "Dubna" has been deployed. During the winter expeditions 2017 and 2018 the number of clusters has been increased by one cluster each year. In the subsequent years during the winter expeditions two more clusters have been deployed consecutively. In 2021 only one cluster has been installed, thus the total number of clusters is recently eight. Thus, the Baikal-GVD telescope comprises $2\,304$ OMs in total. 

Apart of Cherenkov radiation, also the ambient background light is registered. The trigger system of every cluster is designed in such a way that signals from each OM in a time window of 5 $\mu s$ are stored, if a trigger condition is fulfilled. The trigger fires when any pair of neighbor OMs generates signals exceeding low ($\sim$ 1.5 p.e.) and high ($\sim$ 4 p.e.) charge thresholds in a time window less than 100 nano-seconds \cite{trigger2019}. The main source of events passing this trigger are atmospheric muons and random noise. An acquisition of signals from background light is achieved by selecting signals in the first two micro-seconds within the whole 5 $\mu s$ time window. Data on random noise are collected in this way practically continuously as far as the typical trigger rate of a cluster reaches $\sim$ tens of Hz. The amount of the registered background light is derived from the photo-multiplier tube noise rates from each particular OM. The origin of the background noise rates is mainly associated with the luminescence of Baikal water. 

In this article, we present some selected results on luminescence in Lake Baikal observed with Baikal-GVD neutrino telescope.

\section{Ambient light field of Baikal water}

\begin{figure}[h]
\begin{center}$
\begin{array}{cc}
\includegraphics[width=64mm]{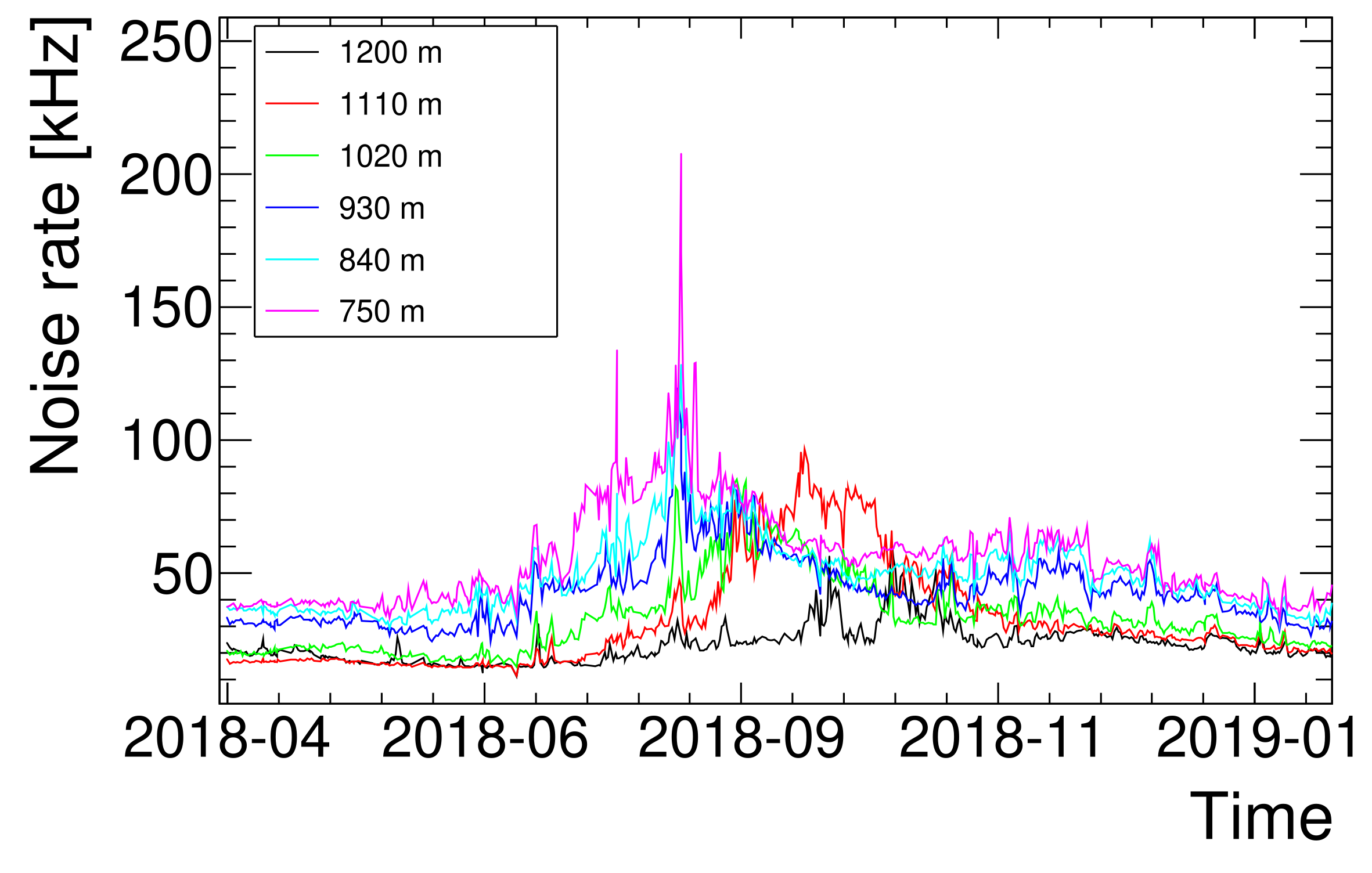}  &
\includegraphics[width=70mm,height=41mm]{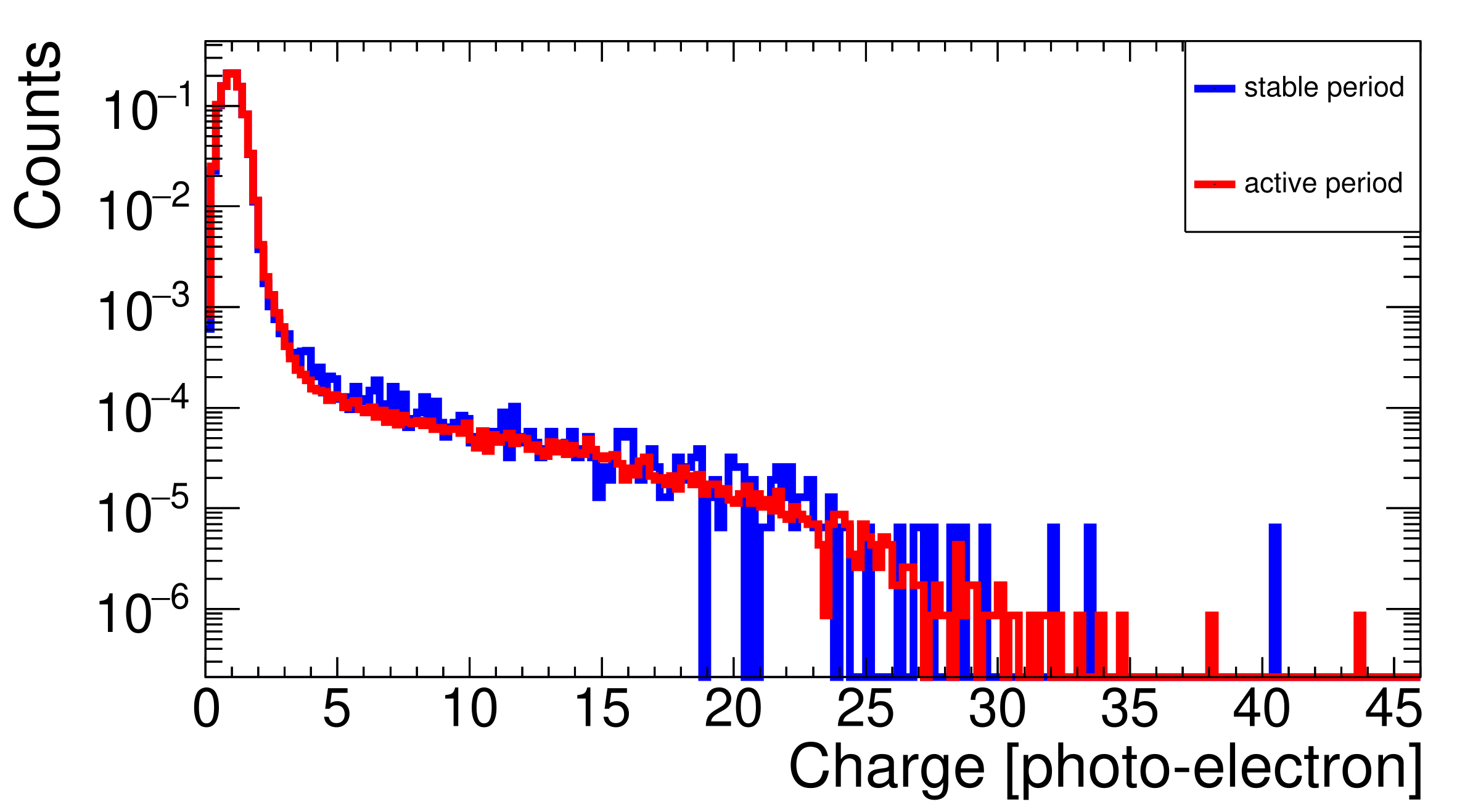}
\end{array}$
\end{center}
\caption{Left panel: Noise rates evolution for OMs at the same string at different depths. There are shown only 6 of 36 OMs for the sake of simplicity placed at depths of 750, 840, 930, 1\,020, 1\,110, and 1\,200 meters. Data from April 2018 till January 2019. Right panel: Charge distribution of registered pulses in units of photo-electrons for the same OM in different periods of optical activity normalized to unity. 
\label{fig.1}}
\end{figure}

The Baikal-GVD neutrino telescope is designed to detect the Cherenkov light from charged particles. In open water, light not related to relativistic particles constitutes an unavoidable background to the Cherenkov light. Therefore studies of the related light fields are of crucial importance. 
An overview of collected data from the the photo-multiplier tube noise rates for selected OMs in different depths for April 2018 -- January 2019 are presented in Fig.\ref{fig.1} (left panel).

There are two periods of relatively stable optical background noise, which are intermitted by increased optical activity. The charge distribution of the noise pulses is displayed in Fig.\ref{fig.1} (right panel). We stress that the charge distribution remains unchanged in different periods of the optical activity. Our measurements are performed with a threshold of one fifth of a single photo-electron charge. In this way, the dark noise of the photo-multiplier tube is significantly suppressed. We note that by setting the threshold to one photo-electron the background count rate is reduced by a factor of two. The one photo-electron background is well correlated with the half photo-electron background. The count rates in both cases exhibit the same modulation of the relative amplitude. We clearly see that the major contribution comes from single photo-electron pulses. 

\begin{figure}[h]
\begin{center}$
\begin{array}{cc}
\includegraphics[width=65mm]{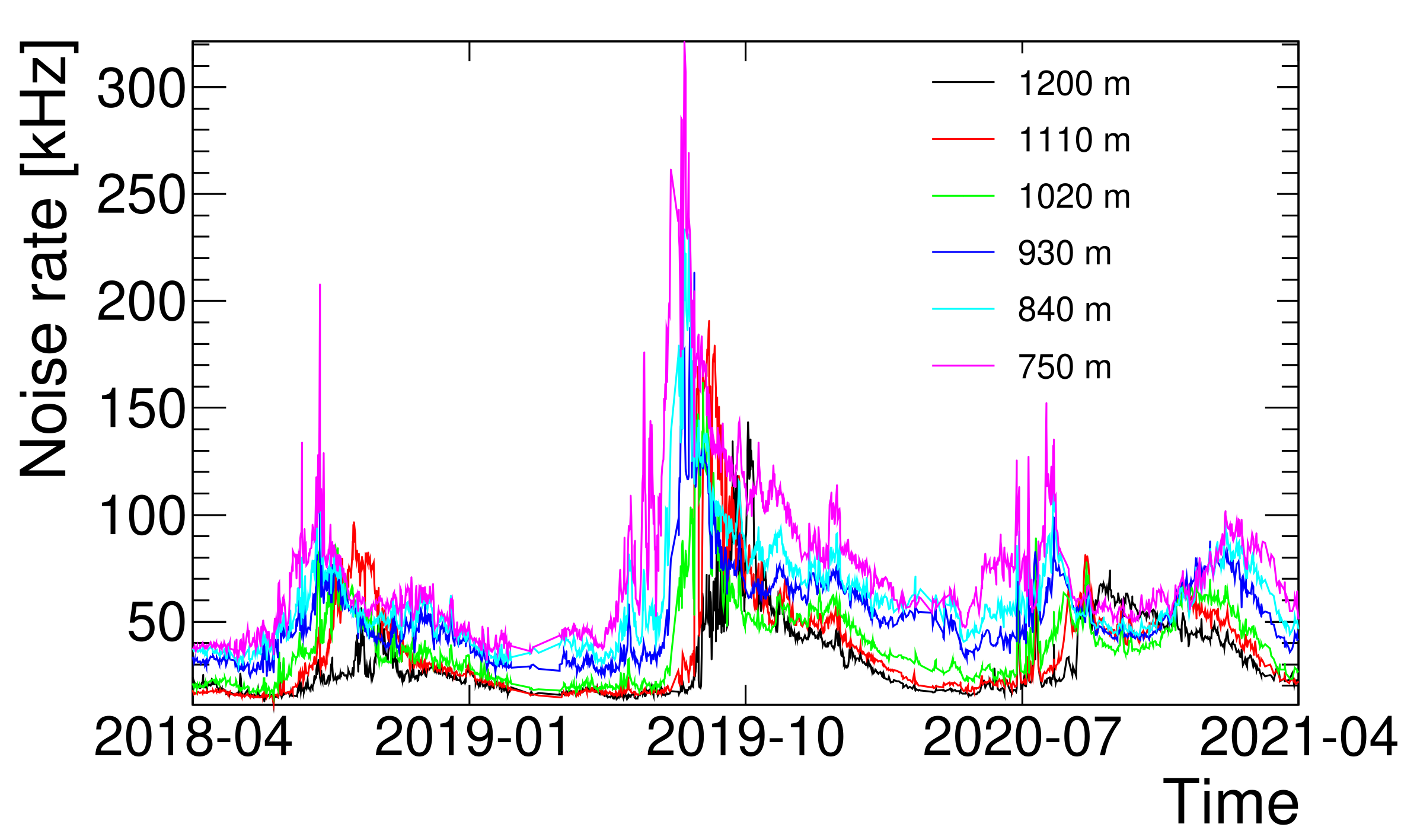}  &
\includegraphics[width=62mm]{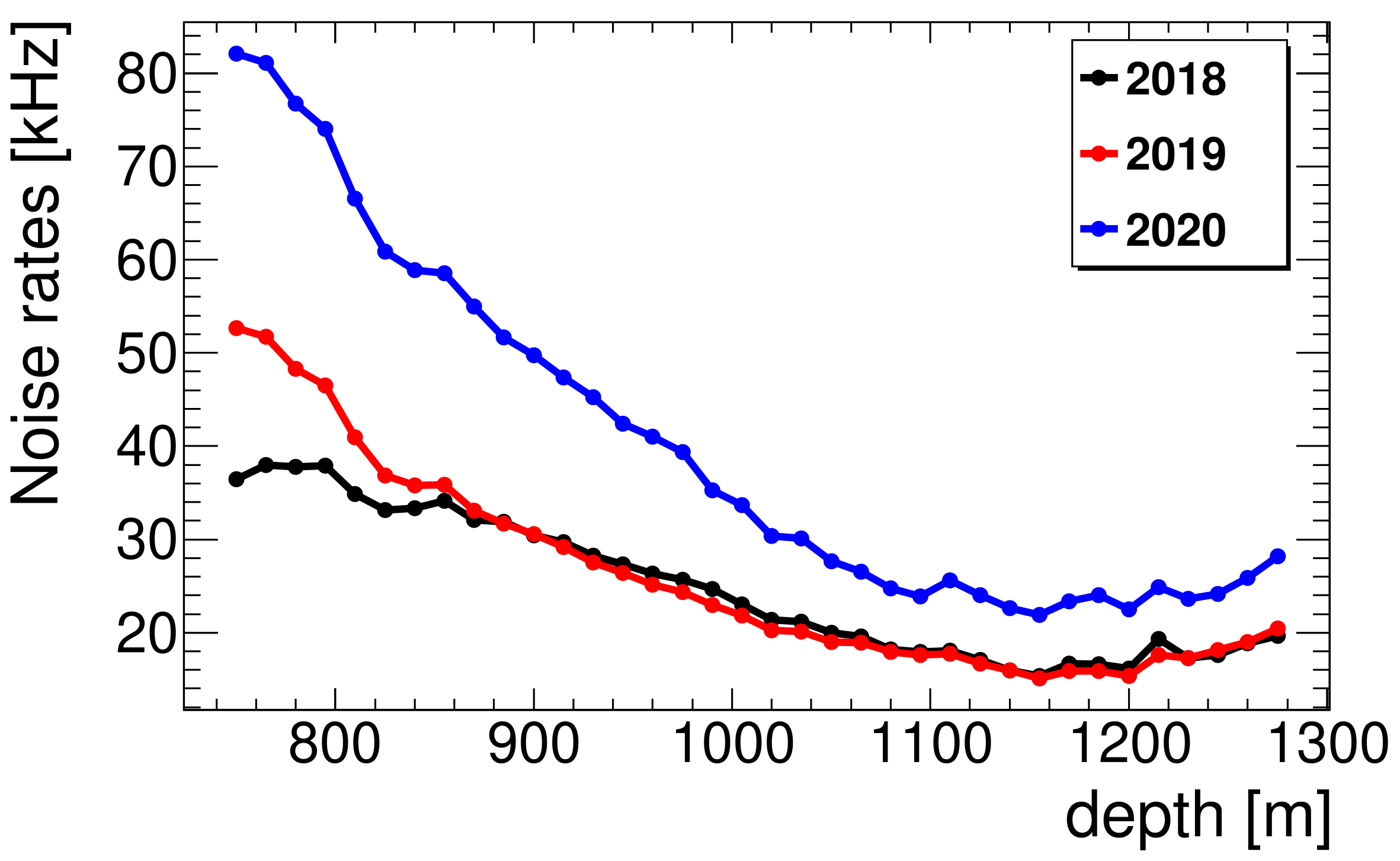}
\end{array}$
\end{center}
\caption{Left panel: Noise rates evolution for OMs at the same string at different depths. Data from April 2018 till April 2021. Right panel: Noise rates as a function of depth averaged over 8 strings of a cluster. Data are from the stable periods from each of the years 2018, 2019, and 2020. The lake bed is at 1\,366 meters depth.
\label{fig.2}}
\end{figure}

\begin{figure}[h]
\begin{center}$
\begin{array}{cc}
\includegraphics[width=70mm]{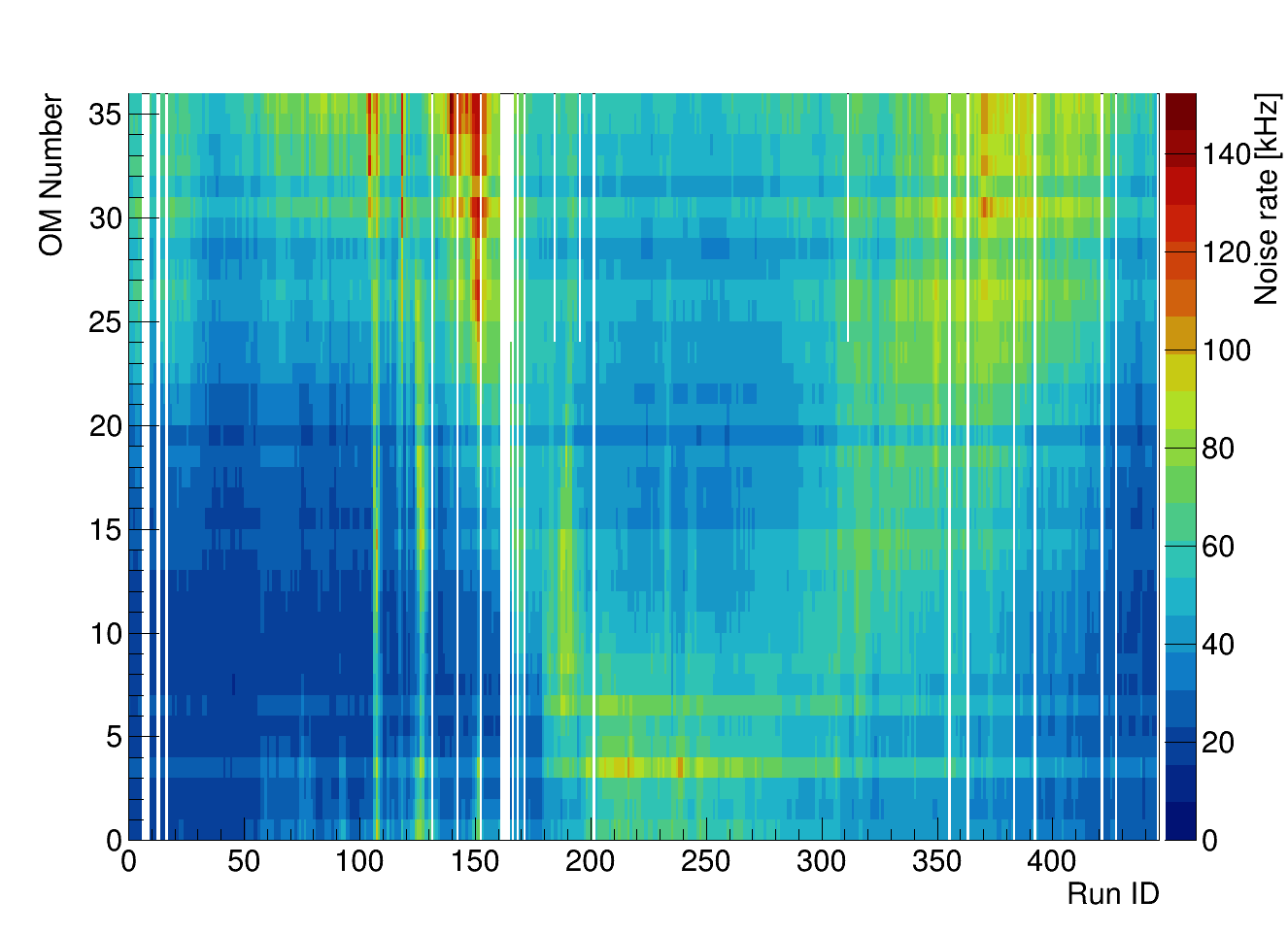}  &
\includegraphics[width=75mm]{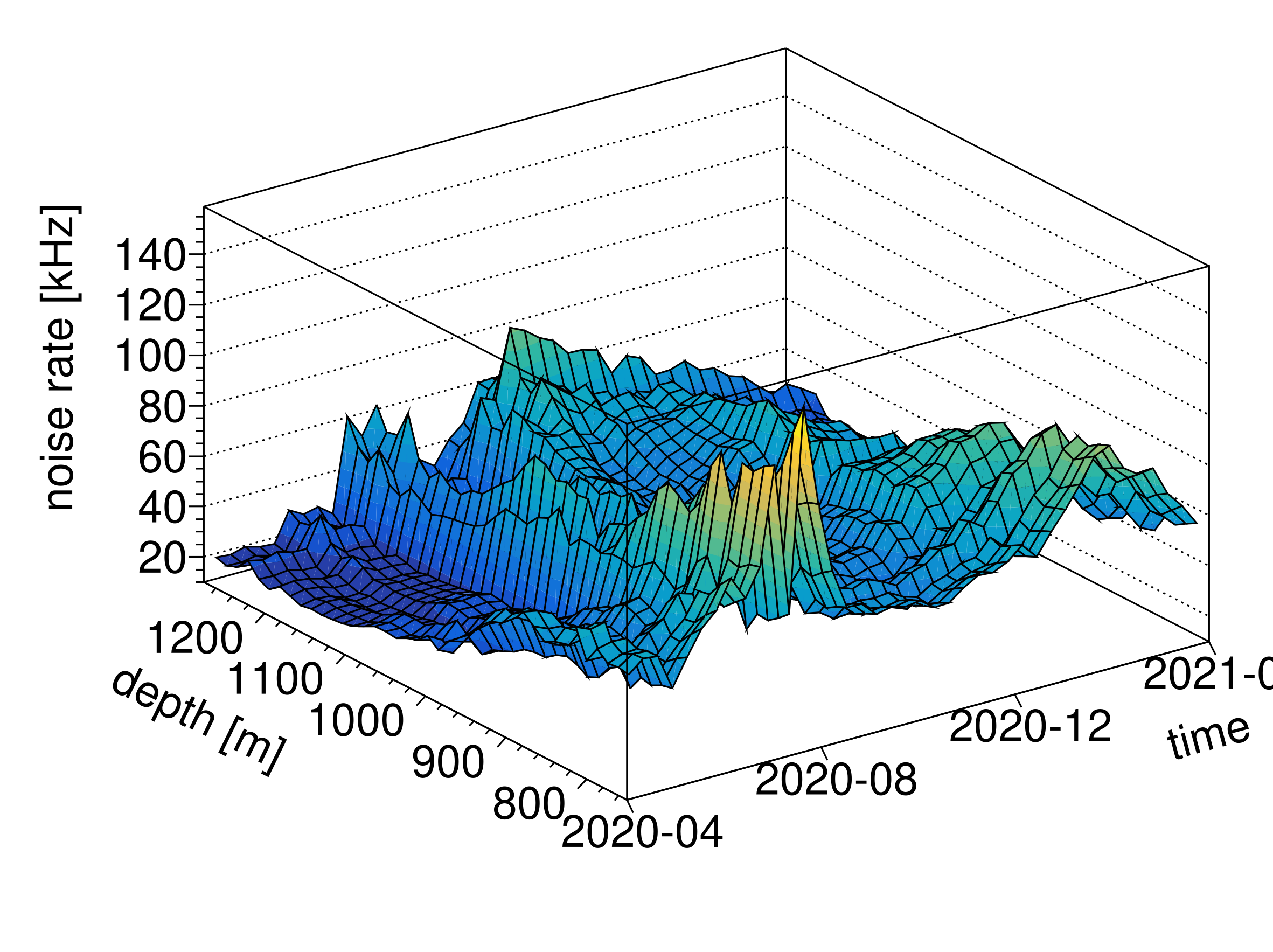}
\end{array}$
\end{center}
\caption{Left panel: Noise rates versus run number are shown for string No.3 of cluster No.3 for the year 2020. OM number 0 and 36 are for the bottom and the top OM. Run ID coincides with the measurement time. Right panel: The same as in the left panel only with OM number replaced with depth in meters and run ID with unix time.
\label{fig.3}}
\end{figure}

The data acquisition is almost continuous without interputions for some clusters of the Baikal-GVD telescope. For instance, the time evolution of the noise rates for the cluster No.3 is displayed in Fig. \ref{fig.2} (left panel). The appearance of the outbreak maximum depends on time, starting with the top modules. Indeed, we observe a layer of highly luminescent water moving from the top to the bottom of the lake every year. By comparing the maximum for different depths, we obtain a velocity profile of the vertical water propagation. In the beginning of August 2018, the estimated speed reached its maximal value of $\sim$16\,m/day, while it remained almost constant ($\sim$5\,m/day) till the end of September 2018, i.e. when the activity asymptotically reached the background plateau. The observed pattern is similar to previous investigations with NT200 detector (see \cite{nonstationarity}). Highly luminescent layer moving upwards has been observed in the year 2021 for the first time, see Fig.\ref{fig.3}. The estimated speed reached its maximal value of $\sim$28\,m/day.

In general, the depth profile of the ambient light field is the same for all eight strings of a cluster. By averaging the count rates over the OMs at the same horizon, we obtain the depth dependence of the background light noise. The average count rates versus the depth are presented in Fig.\ref{fig.2} (rigt panel). They are taken from the quite periods for each of the years 2018, 2019, and 2020. The photon flux from the sunlight below a depth of $\sim$ 700 m is negligible as shown in previous work \cite{luminescence98}. The analysed data are from periods of stable noise activity in years 2018, 2019, and 2020 for the same cluster. The pattern remains the same for other clusters. An analytical approximation of the depth dependence, which takes the form
\begin{eqnarray}
f(H)&\approx& \exp^{-\frac{H}{H_0}},
\end{eqnarray}
was presented in the previous work \cite{budnev2018}. $f$ and $H$ are average noise rates in kHz and depth in meters, respectively. $H_0$ is the free parameter with fit results of the experimental data equal to $461,~312$, and $292$ meters for the seasons 2018, 2019, and 2020, respectively. 

\begin{figure}[h]
\begin{center}$
\begin{array}{cc}
\includegraphics[width=68mm]{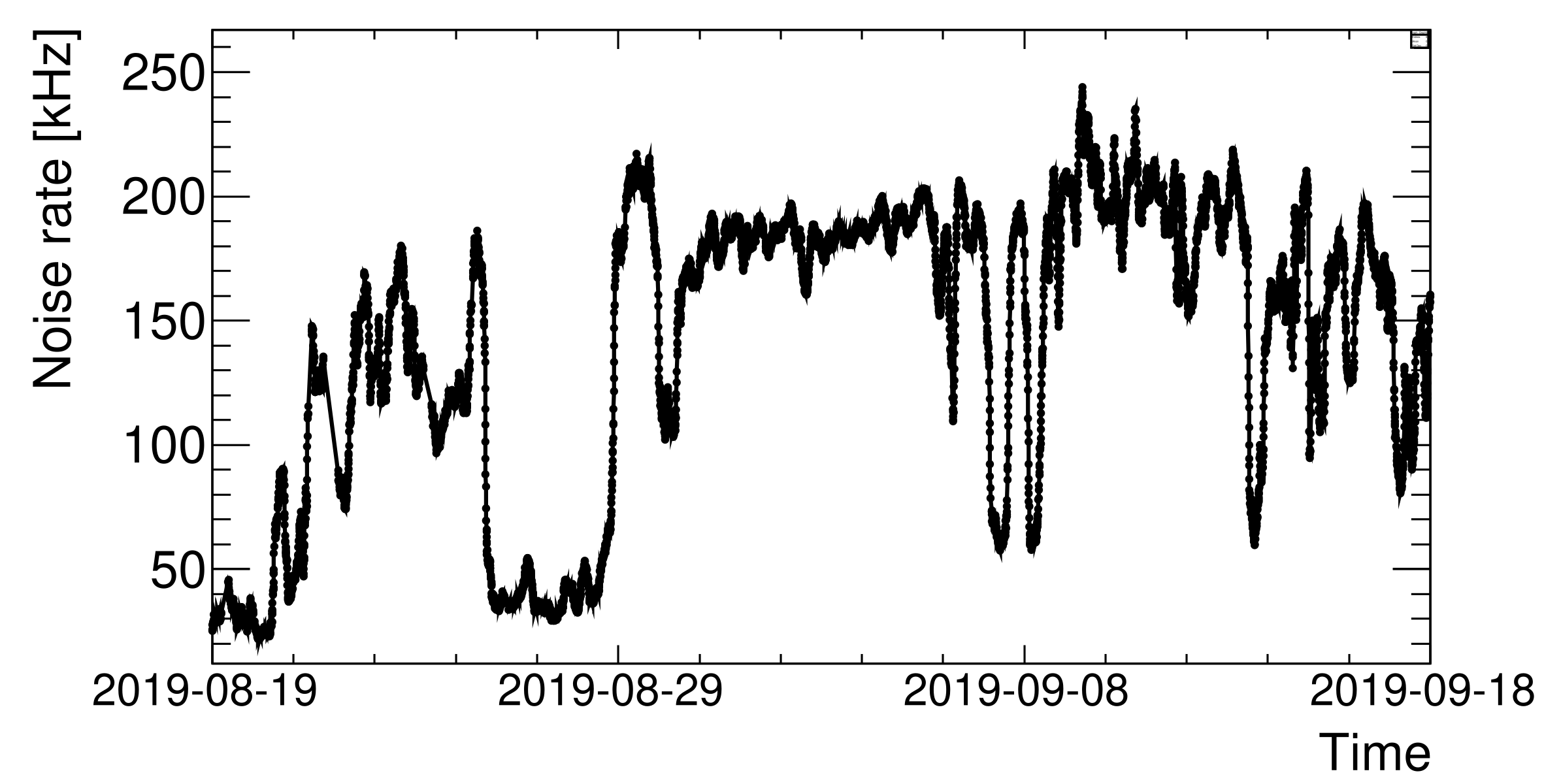}  &
\includegraphics[width=65mm,height=41mm]{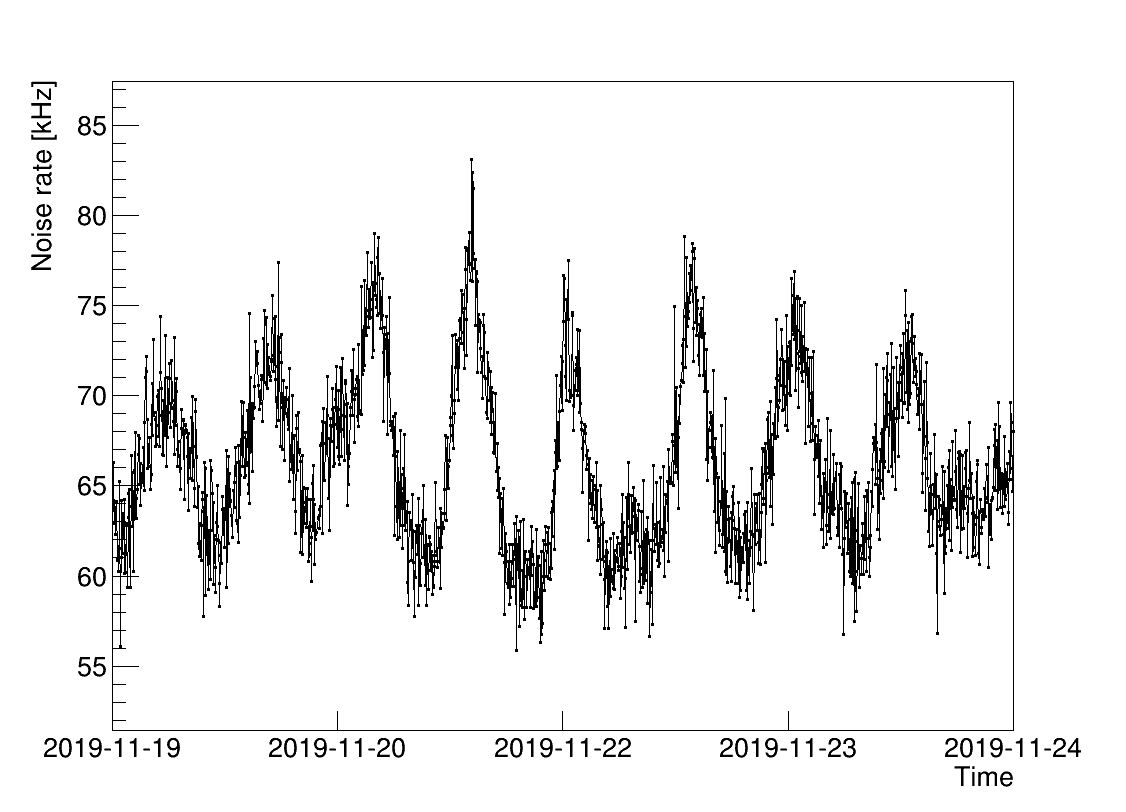}
\end{array}$
\end{center}
\caption{Left panel: Noise rates for a particular OM during the optically highly active period, namely from August till September 2019. A sudden outburst of the count rates is noticeable. Right panel: An example of the regular modulation of noise rates. Data are from the period of stable plateau, namely from November 2019 till.   
\label{fig.4}}
\end{figure}

The time evolution of the noise rates, as shown in Fig.\ref{fig.2} (left panel), exhibits sharp changes on top of relatively continuous smooth optical background. The effect is more visible in particularly selected time window displayed in Fig. \ref{fig.4} (left panel). The amplitude of these sudden changes reached almost 150\,kHz. The duration of such variations which distort the smooth background ranges typically from several hours up to a few days. We note that effect is present in July -- September almost every year, i.e. the period of increased luminescent activity. However, the period of relatively stable plateau (for instance October -- February 2019) shows (see Fig. \ref{fig.4}, right panel) regular modulation of noise rates. The period of these modulations is quite stable and varies from 10 to 14 hours. 

Due to the horizontal water currents in Lake Baikal, the strings geometry deviates from its vertical direction. To take these deviations into account, an acoustic positioning system for Baikal-GVD has been developed  \cite{acou2019}. Torrent flows in the lake, appearing during the autumn period of the year, may produce a remarkable tilt of the string from its vertical position. Our observations do not find a correlation between the torrent string deviations and the luminescence activity of the lake, which is in an agreement with our previous studies \cite{lumi2019}.

\section{Conclusions}

We have presented data on the luminescence in Lake Baikal which have been collected by the Baikal-GVD neutrino telescope. We found increases of the luminescence activity intermitting periods of relatively stable optical background in 2018, 2019, and 2020. For the first time we found that the maximum of the optical activity observed in January 2021 propagated from bottom to the top, with a maximum speed of 28\,m/day. The maximal amplitude of the sudden modulations in an active period reached almost 150\,kHz. 

\section{Acknowledgements}

The work was partially supported by RFBR grant 20-02-00400. The CTU group acknowledges the support by European Regional Development Fund-Project No. CZ.02.1.01/0.0/0.0/16\_019/0000766. The CU group acknowledges the support by the VEGA Grant Agency of the Slovak Republic under Contract No. 1/0607/20. We also acknowledge the technical support of JINR staff for the computing facilities (JINR cloud).

\end{document}